\documentclass[11pt]{article}
\usepackage{amsmath,amssymb,amsthm,amsxtra,overpic,bbm,bm,epsfig}
\usepackage{color,subcaption,longtable,blindtext,lscape,caption}
\usepackage{soul}
\usepackage[dvipsnames]{xcolor}

\def\thefootnote{\fnsymbol{footnote}}
\allowdisplaybreaks[4]

\newcommand{\bq}{\begin{eqnarray}}
\newcommand{\nq}{\end{eqnarray}}

\begin{document}

\begin{center}
{\Large\bf Effective alignments and the landscape of $S_4$ flavour models} \\

\end{center} 
\vspace{0.2cm}

\begin{center}
{\bf Ivo de Medeiros Varzielas$^1$}\footnote{Email: \tt ivo.de@udo.edu}
,
{\bf Miguel Levy$^1$}\footnote{Email: \tt plevymiguel@gmail.com}
and 
{\bf Ye-Ling Zhou$^2$}\footnote{Email: \tt ye-ling.zhou@soton.ac.uk}
\\\vspace{5mm}
{$^1$CFTP, Departamento de F\'{\i}sica, Instituto Superior T\'{e}cnico,}\\
Universidade de Lisboa,
Avenida Rovisco Pais 1, 1049 Lisboa, Portugal \\
{$^2$ School of Physics and Astronomy, University of Southampton,\\
Southampton SO17 1BJ, United Kingdom } 
\\
\end{center}

\vspace{1.5cm} 

\begin{abstract} 
We explore the concept of effective alignments: contractions of multiple flavour symmetry breaking flavon fields. These contractions give rise to directions that are hard or impossible to obtain directly by breaking the flavour symmetry. Within this context, and using $S_4$ as the flavour symmetry to exemplify, we perform a phenomenological check of lepton flavour models built from pairing any two effective alignments up to order 2 (in flavon contractions). The check is performed for each pair of effective alignments in a framework with models of constrained sequential dominance type, in a basis where the charged leptons are diagonal. We thus obtain an indication of which effective alignments are interesting for model building, within this so-called $S_4$ landscape. We find three types of viable topologies and provide examples of models realizing this strategy for each topology.
\end{abstract}
\begin{flushleft}
\hspace{0.8cm} PACS number(s): 14.60.Pq, 11.30.Hv, 12.60.Fr \\
\hspace{0.8cm} Keywords: Lepton flavour mixing, flavour symmetry
\end{flushleft}

\def\thefootnote{\arabic{footnote}}
\setcounter{footnote}{0}

\newpage


\section{Introduction}

The formulation of the Standard Model (SM) was one of the biggest successes of particle physics, accounting for most interactions of matter known to date. Notwithstanding its success, it is still a theory that leaves unsanswered some questions. Why are there three copies of fermions? Why are their masses hierarchical? What gives rise to the specific mixing patterns observed? These questions (among others) are part of what is known as the \emph{flavour problem} \cite{King:2014nza}.

The inclusion of a flavour symmetry to the SM became a prolific way to attempt to meaningfully answer some of the questions posed by the flavour problem. These flavour symmetries affect not the gauge structure of the model, rather they restrict the way the different particles are able to interact with each other in the Yukawa sector. As such, this strategy has been widely used to predict the leptonic mixing structure, for which the flavour symmetries favoured are typically non-Abelian discrete symmetries. In these flavour models, we can take the SM to be a low-energy version of a more complete model which includes a flavour symmetry that has been broken by either scalar fields that are singlets with respect to the SM gauge structure (these are called flavons and it is this paradigm we will be discussing), or by enlarging the Higgs sector.
Then, the vacuum expectation values (VEV) of the scalars dictate the Yukawa structures. In leptonic flavour models, the type-I seesaw mechanism is employed and Sequential Dominance (SD) \cite{King:1998jw, King:1999cm, King:1999mb, King:2002nf} is a useful paradigm for model building, with Constrained Sequential Dominance (CSD)
\cite{King:2005bj} remaining a viable strategy to produce predictive models \cite{Antusch:2011ic, King:2013xba, Bjorkeroth:2014vha, Bjorkeroth:2015ora, Bjorkeroth:2015uou, Bjorkeroth:2017ybg}.

Since in the indirect approach models the mixing structure is defined by the flavon VEVs, the problem trickles down to choosing appropriate VEVs to match the observed mixing pattern. Nevertheless, this is still a question that must be in accordance to the scalar potential of the chosen flavour model (both the flavour symmetry chosen and the number of flavons will influence the outcome). Moreover, cross-terms in the scalar potential that arise from the interaction of the different flavons will, in general, influence the direction of the flavon VEVs \cite{Pascoli:2016eld}.

One novel way to explore possible leptonic flavour models is through effective alignments (EAs) \cite{deMedeirosVarzielas:2017hen}. EAs are directions that arise from higher-order operators in which multiple flavons contract, giving rise to a much richer structure. In sum, starting from simple flavon VEVs, we can achieve more complex EAs. In this paper, we will focus on the $S_4$ symmetry group, a group with a long tradition of being used to explain large neutrino mixing angles \cite{Mohapatra:2003tw}. We consider contractions of no more than two $S_4$ flavons, and do an exploratory survey of the resulting landscape, to ascertain at a primary level if there are any EAs compatible with current experimental data of neutrino oscillations.

After reviewing the list of possible minima of the $S_4$ scalar potential \cite{Ivanov:2014doa, deMedeirosVarzielas:2017glw, deMedeirosVarzielas:2017ote}, we take these to be the possible VEV alignments our flavons can take. These will be the starting point from which EAs can be constructed, through contractions up to two flavons.

The $S_4$ symmetry group, this mechanism of flavon contraction, as well as the list of possible VEV alignments, will be the focus of section \ref{ssec:vev conf}. Afterwards, in section \ref{sec:lep_mix}, we consider the leptonic sector of CSD models, describing how lepton mixing is formulated in this framework, as well as how pairs of EAs can be used to produce leptonic mixing. In section \ref{sec:landscape}, we go through the numerical method used, and present the obtained results, to ascertain which of the possible EAs studied could lead to the observed leptonic mixing. In Section \ref{sec:models}, we present specific UV complete \cite{Varzielas:2010mp, Varzielas:2012ai, Ding:2013hpa, Ding:2014hva} realizations of models with the pairs of EAs that were found to lead to viable mixing. Finally, we present our conclusions in section \ref{sec:Conclusions}.

\section{Flavon vacuum alignments in $S_4$ \label{sec:VEV}}

We briefly review the idea of effective alignments (EAs) \cite{deMedeirosVarzielas:2017hen} in the flavour space.

The theory is assumed to be invariant under a flavour symmetry. As in \cite{deMedeirosVarzielas:2017hen}, we consider as example  $S_4$, the group of the permutations of four objects, see e.g.,\ \cite{Escobar:2008vc, Ding:2013eca, Li:2014eia}. It has 24 distinct elements,
which are generated by three generators $S$, $T$ and $U$ satisfying
the relations
$T^3=S^2=U^2=(ST)^3=(SU)^2=(TU)^2=1$,
which automatically leads to $(STU)^4=1$. 
$S_4$ has 5 irreducible representations (irreps), $\mathbf{1}$, $\mathbf{1}'$, $\mathbf{2}$, $\mathbf{3}$ and $\mathbf{3}$. Representation matrices of generators are given in Table \ref{tab:rep_matrix}, Appendix \ref{app:S4}. 

\subsection{Flavon VEV configurations \label{ssec:vev conf} }

Following the discussion in \cite{deMedeirosVarzielas:2017hen}, we take the $S_4$ potential of a complex triplet flavon $\varphi=(\varphi_1, \varphi_2, \varphi_3)^T$, in the irreducible representation $\mathbf{3}$ or $\mathbf{3^\prime}$ of $S_4$.
By embedding an additional $U(1)$ symmetry, the renormalisable flavon potential takes a simple form as
\bq
V(\varphi)&=& \mu^2_\varphi I_1 + g_1 I_1^2 + g_2 I_2 + \frac{1}{2}g_3( I_3 + \text{h.c.})
\label{eq:Vphi}
\nq
with
\bq
I_1 &=& |\varphi_1|^2 + |\varphi_2|^2 + |\varphi_3|^2\,,\nonumber\\
I_2 &=& |\varphi_1\varphi_2|^2 + |\varphi_2\varphi_3|^2 + |\varphi_3\varphi_1|^2\,,\nonumber\\
I_3 &=& (\varphi_1^*\varphi_2)^2 + (\varphi_2^*\varphi_3)^2 + (\varphi_3^*\varphi_1)^2\,.
\nq
Vacuum stability requires the conditions $\mu_\varphi^2<0$, $g_1>0$ and $3g_1+g_2>|g_3|$ to be satisfied at tree level.

The vacuum is selected by minimising the potential, in detail, by solving the extremisation condition $\partial V(\varphi)/\partial \varphi_i = 0$ and requiring the positivity condition $\partial^2 V(\varphi)/\partial \varphi_i \partial \varphi_j$. 
The equation $\partial V(\varphi)/\partial \varphi_i = 0$ can be solved explicitly. Up to a normalisation factor, there are in total 21 VEV candidates obtained \cite{deMedeirosVarzielas:2017hen}. They are classified into 4 classes, which are given in the basis of \cite{Altarelli:2005yp, Altarelli:2005yx} by
\bq
a &=& \frac{1}{\sqrt{3}}\left\{ \hspace{5mm}
\begin{pmatrix}1 \\ 1 \\ 1 \end{pmatrix} \,,\hspace{23mm}
\begin{pmatrix}1 \\ \omega^2 \\ \omega \end{pmatrix} \,,\hspace{22mm}
\begin{pmatrix}1 \\ \omega \\ \omega^2 \end{pmatrix}  \hspace{5mm}
\right\}\,;\nonumber\\
&& \hspace{18mm} a_1
\hspace{32mm} a_2
\hspace{32mm} a_3 \nonumber\\
b &=& \left\{ 
\begin{pmatrix} \frac{1}{\sqrt{3}} \\ \frac{1}{3+\sqrt{3}} \\ \frac{-1}{3-\sqrt{3}} \end{pmatrix}, 
\begin{pmatrix} \frac{1}{\sqrt{3}} \\ \frac{-1}{3-\sqrt{3}} \\ \frac{1}{3+\sqrt{3}} \end{pmatrix}, 
\begin{pmatrix} \frac{1}{\sqrt{3}} \\ \frac{\omega^2}{3+\sqrt{3}} \\ \frac{-\omega }{3-\sqrt{3}} \end{pmatrix}, 
\begin{pmatrix} \frac{1}{\sqrt{3}} \\ \frac{-\omega^2}{3-\sqrt{3}} \\ \frac{\omega }{3+\sqrt{3}} \end{pmatrix}, 
\begin{pmatrix} \frac{1}{\sqrt{3}} \\ \frac{\omega }{3+\sqrt{3}} \\ \frac{-\omega^2}{3-\sqrt{3}} \end{pmatrix}, 
\begin{pmatrix} \frac{1}{\sqrt{3}} \\ \frac{-\omega }{3-\sqrt{3}} \\ \frac{\omega^2}{3+\sqrt{3}} \end{pmatrix} 
\right\} \,; \nonumber\\
&& \hspace{10mm} b_1
\hspace{14mm} b_1^*
\hspace{14mm} b_2
\hspace{15mm} b_2^*
\hspace{14mm} b_3
\hspace{15mm} b_3^* \nonumber\\
c &=& \left\{ \hspace{5mm}
\begin{pmatrix} \frac{1}{3} \\ -\frac{2}{3} \\ -\frac{2}{3} \end{pmatrix}, \hspace{18mm}
\begin{pmatrix} \frac{1}{3} \\ -\frac{2}{3}\omega^2 \\ -\frac{2}{3}\omega \end{pmatrix}, \hspace{18mm}
\begin{pmatrix} \frac{1}{3} \\ -\frac{2}{3}\omega \\ -\frac{2}{3}\omega^2 \end{pmatrix}, \hspace{15mm}
\begin{pmatrix} 1 \\ 0 \\0 \end{pmatrix}
\right\} \,; \nonumber\\
&& \hspace{13mm} c_1
\hspace{31mm} c_2
\hspace{33mm} c_3
\hspace{27mm} c_4 \nonumber\\
d &=& \left\{ 
\begin{pmatrix} \frac{2}{3} \\ \frac{2}{3} \\ -\frac{1}{3} \end{pmatrix},
\begin{pmatrix} \frac{2}{3} \\ -\frac{1}{3} \\ \frac{2}{3} \end{pmatrix},
\begin{pmatrix} -\frac{2}{3}\omega \\ -\frac{2}{3}\omega^2 \\ \frac{1}{3} \end{pmatrix},
\begin{pmatrix} -\frac{2}{3}\omega^2 \\ \frac{1}{3} \\ -\frac{2}{3}\omega \end{pmatrix},
\begin{pmatrix} -\frac{2}{3}\omega^2 \\ -\frac{2}{3}\omega \\ \frac{1}{3} \end{pmatrix},
\begin{pmatrix} -\frac{2}{3}\omega \\ \frac{1}{3} \\ -\frac{2}{3}\omega^2 \end{pmatrix},
\begin{pmatrix} 0 \\ 0 \\ 1 \end{pmatrix},
\begin{pmatrix} 0 \\ 1 \\ 0 \end{pmatrix}
\right\} \,, \nonumber\\
&& \hspace{8mm} d_1
\hspace{10mm} d_1^*
\hspace{14mm} d_2
\hspace{14mm} d_2^*
\hspace{15mm} d_3
\hspace{15mm} d_3^*
\hspace{11mm} d_4
\hspace{7mm} d_4^* 
\label{eq:flavon_vevs}
\nq
where $\omega=e^{i2\pi/3}$. Whether these solutions are global minimums depends on values of $g_1$, $g_2$ and $g_3$.
\begin{itemize}
\item For $g_2 + g_3 > 0$  and $ g_2-g_3 >0$, class $a$ is the global minimum. 

\item For $g_2 + g_3 > 0$  and $ g_2-g_3 <0$, class $b$ is the global minimum. 

\item For $g_2 + g_3 <0$ and $g_3 < 0$, class $c$ is the global minimum.

\item For $g_2 + g_3 >0$ and $g_3 > 0$, class $d$ is the global minimum.
\end{itemize}

All solutions in each class (e.g., $a_1$, $a_2$ and $a_3$ in class $a$) are degenerate. In the single flavon case, it is a random choice to select any of them. However, once the model involves several flavons whose VEVs are in the same class, the situation with VEVs taking the same solution is physically different from the case where they take different solutions \cite{King:2019tbt}. Furthermore, there may be some residual symmetry preserved in some of these VEVs, e.g., $a_1$ preserving $Z_2=\{1,S\}$, $c_4$ preserving $Z_3=\{1,T,T^2\}$. For more discussion on this aspect, please see \cite{deMedeirosVarzielas:2017hen}. 

We collect all these solutions in the set 
\bq
S^{(1)}_{\text{VEV}}= \{ a_1, a_2, a_3, b_1, b_1^*, b_2, b_2^*, b_3, b_3^*, c_1, c_2, c_3, c_4, d_1, d_1^*, d_2, d_2^*, d_3, d_3^*, d_4, d_4^* \} \,.
\label{eq:S1VEV}
\nq 
In the following, we regard the 21 VEV alignments in $S^{(1)}_{\text{VEV}}$ as building blocks of the other EAs. 
For any $x, y, z, \cdots \in S^{(1)}_{\text{VEV}}$, the 3-dimensional irreducible representations $(xy)_{\mathbf{3}}$ and $(xy)_{\mathbf{3}'}$ form a set $S^{(2)}_{\text{VEV}}$, and $((xy)_{\mathbf{3}}z)_{\mathbf{3}}$, $((xy)_{\mathbf{3}}z)_{\mathbf{3}'}$, $((xy)_{\mathbf{3}'}z)_{\mathbf{3}}$ and $((xy)_{\mathbf{3}'}z)_{\mathbf{3}'}$ form another set $S^{(3)}_{\text{VEV}}$, and so on. Following the procedure, one can in principle get $S^{(3)}_{\text{VEV}}$, $S^{(4)}_{\text{VEV}}$, .... We define their connection as $S_{\text{VEV}}$, 
\bq
S_{\text{VEV}} = S^{(1)}_{\text{VEV}} \bigcup S^{(2)}_{\text{VEV}} \bigcup S^{(3)}_{\text{VEV}} \bigcup \cdots\,.
\label{eq:vev union}
\nq 
We have checked that from the 21 EAs in $S^{(1)}_{\text{VEV}}$, 174 additional EAs are generated in $S^{(2)}_{\text{VEV}}$ and another 2988 EAs are generated in $S^{(3)}_{\text{VEV}}$. 

\subsection{Lepton mixing in the CSD framework \label{sec:lep_mix}} 

We can now use the flavon VEVs and the EAs to build models in the framework of CSD. We write the Lagrangian terms as
\begin{eqnarray}
-\mathcal{L}_l &=& (Y_\tau' \overline{\ell_L})_{\mathbf{1'}} H \tau_R 
+ \big( Y_\mu \overline{\ell_L}\big)_{\mathbf{1}} H \mu_R  
+(Y_e' \overline{\ell_L} )_{\mathbf{1'}} H e_R 
+ \text{h.c.} \,, \nonumber\\
-\mathcal{L}_\nu &=& \sum_{i}\, (Y_i^{(\prime)} \overline{\ell_L})_{\mathbf{1}^{(\prime)}} \tilde{H} N_i + \frac{1}{2} M_i\overline{N^c_i} N_i + \text{h.c.} \,, 
\label{eq:L_N}
\end{eqnarray} 
where $i$ labels the different right-handed neutrinos (RHNs). The minimal case corresponding to a massless active neutrino occurs for $i=1,2$. We assume for this paper that the charged lepton Yukawa couplings are diagonal up to a row permutation, namely that 
\bq \label{eq:lepton_VEV}
Y_{e,\mu,\tau}\propto \begin{pmatrix} 1 \\ 0 \\ 0 \end{pmatrix}\,,~\begin{pmatrix} 0 \\ 1 \\ 0 \end{pmatrix}\,,~\begin{pmatrix} 0 \\ 0 \\ 1 \end{pmatrix}\,.
\nq 
This is easily achieved with $S_4$ in the basis we are working on, e.g. by using the VEV $d_4=(0,0,1)^T$ and EAs built from it (e.g., in \cite{deMedeirosVarzielas:2017hen} the flavon $\Phi_{l'}$ had the VEV of this form). Other EAs, e.g., $c_4=(1,0,0)^T$, $d_4^*=(0,1,0)^T$ or their contractions can also be used to construct a diagonal charged lepton mass matrix up to a row permutation.
With this assumption, the model building basis coincides with the basis where charged leptons are diagonal and mixing comes just from the neutrino sector, where the mass matrix is
\bq
M_\nu &=& \sum_{i}\, \frac{v^2}{M_{N_i}} Y_i^{(\prime)} Y_i^{(\prime)T}  
= \sum_{i} \mu_i V_i V_i^T \,,
\label{eq:neutrino_mass}
\nq
$\mu_i$ being complex mass parameters. Here, $V_i$ and all EAs in Eq.~\eqref{eq:lepton_VEV} should be included in the set $S_{\text{VEV}}$. 

We focus on models with only two RHNs ($i=1,2$) which are particularly predictive, one $\mu_i$ vanishes and correspondingly one of the active neutrinos is massless. We can then distinguish two cases
\begin{itemize}
\item $V_1$ is orthogonal to $V_2$, $V_1^\dag V_2=0$. This completely fixes the PMNS mixing matrix 
\bq
U_\text{PMNS}=(U_1, V_1, V_2) \;\;\text{or}\;\; (U_1, V_2, V_1) \,,
\nq 
up to Majorana phases and a row permutation.
The mixing matrix is unrelated to neutrino masses (so-called mass-independent mixing schemes or form-diagonalizable schemes \cite{deMedeirosVarzielas:2011tp}).
To fit current data, often small corrections $\delta V_2$ or $\delta V_3$ are necessary (see e.g.\ \cite{Sierra:2013ypa, Sierra:2014hea}). 
\item $V_1$ is not orthogonal to $V_2$. The mixing matrix can be parametrised by a rotation between the 2nd and 3rd columns,
\bq
U_\text{PMNS}=(U_1, V_1, V_2) \begin{pmatrix} 1 & 0 & 0 \\ 0 & \cos\theta & \sin\theta e^{-i \gamma} \\ 0 & -\sin\theta e^{i \gamma} & \cos\theta \end{pmatrix} \,,
\nq 
$\theta$ and $\gamma$ depend on $\mu_1/\mu_2$, i.e.\ depend on the the neutrino mass ratio. A well known example is TM1 mixing \cite{Xing:2006ms, Lam:2006wm, Albright:2008rp, Albright:2010ap}, when $V_1, V_2 \perp (2, -1, -1)^T$. $S_4$ can be used to achieve TM1 \cite{Varzielas:2012pa}, as can other groups \cite{Varzielas:2015aua}. Most of the models we consider here are examples of this type, as are the CSD2 models with 2 RHN neutrinos \cite{Antusch:2011ic, Bjorkeroth:2014vha}. The ambiguity of the permutation of the rows of the PMNS matrix will always be taken into account.
\end{itemize}

\section{The $S_4$ landscape \label{sec:landscape}}

We are interested in testing the landscape of $S_4$ flavour models in the CSD paradigm described in Section \ref{sec:lep_mix}. We thus test each pair of EAs resultant of Eq. (\ref{eq:vev union}) up to $S^{(2)}_{\text{VEV}}$.
This includes cases with just two flavons (one flavon for each EA), with three flavons (one flavon for one EA and two for the other) and four flavons (two flavons for each EA).

Following the steps in the Appendix of \cite{Antusch:2003kp}, the mixing angles, $\delta_{CP}$, and the two mass-squared differences are extracted from the neutrino mass matrix (we take the ambiguity of the permutation of the rows of the PMNS matrix into account, as discussed in Section \ref{sec:lep_mix}). Then a standard $\chi^2$ analysis is performed to compare these observables and ascertain to which extent each model (corresponding to a pair of EAs) is compatible with experimental data. The $\chi^2$ is computed individually for each experimental value, 
\begin{equation}
\chi^2_i = \frac{(x_i^{\text{model}} - x_i^{\text{bfp}})^2}{\sigma_{x_i}^2}, 
\end{equation}
where $x_i$ is any one of the six test metrics (meaning the six tested observables: three mixing angles, two mass-squared differences, and a phase), $x_i^{\text{model}}$ is the value the test metric assumes for that particular model, $x_i^{\text{bfp}}$ is the best-fit point, and $\sigma_{x_i}$ is the corresponding difference between the three (or one) $\sigma$ values for the test metric. Since the best-fit point is not perfectly centred around the three (one) $\sigma$ regions, the $\sigma_{x_i}$ values are distinct whether the test metric for the model is larger, or smaller, than the best-fit point. A model is then considered compatible with experimental data (taken from \cite{deSalas:2017kay}) when $\chi_ i^2 \leq 1$ for all six test metrics. 

In order to achieve the optimal $x_i^{\text{model}}$ values for the test metrics for each model, we compute the sum over all $\chi_i^2$, which acts as the minimizing value that is the target for the MINUIT minimization routine \footnote{The MINUIT minimization routine is part of the CERN program library \cite{James:2004xla}.}. This routine takes random starting point values for the model's free parameters (the $\mu_i$), and then searches for optimal values for these values which minimize the target variable ( $\sum_i \chi_i^2$ in our case). In this way, the MINUIT routine will search for the optimal set of values of the free parameters space which results in the average best fit for the model's predictions (our six test metrics), given the best-fit points of each test metric and the chosen validity interval (three or one $\sigma$ range).
After searching for EA pairs compatible with the $1\sigma$ experimental range for all observables and not finding any, we always set $\sigma_{x_i}$ to the respective $3 \sigma$ range.

\begin{figure}
\begin{subfigure}[b]{0.5\textwidth}
        \includegraphics[width=1\textwidth,trim={15mm 5mm 25mm 0},clip]{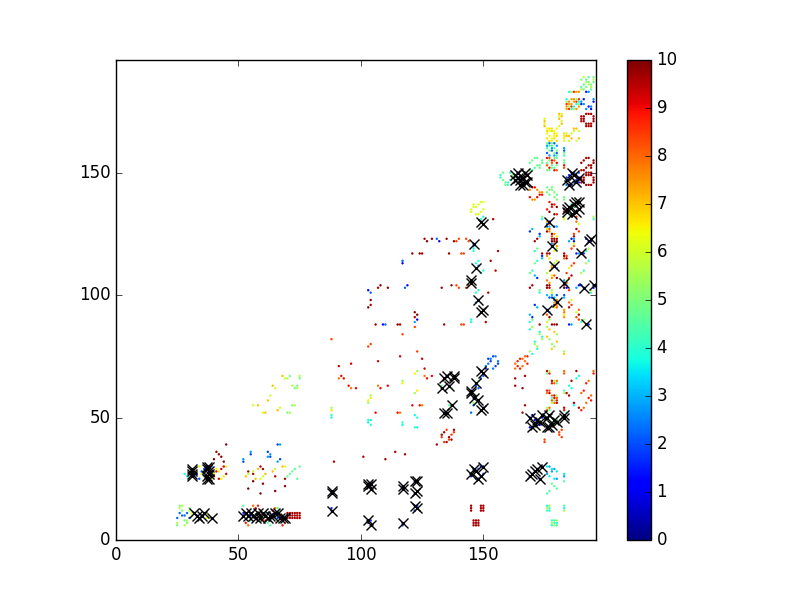}
        \caption{Normal ordering, $\chi^2<10$}
        \label{subfig:a}
\end{subfigure}
\begin{subfigure}[b]{0.5\textwidth}
        \includegraphics[width=1\textwidth,trim={15mm 5mm 25mm 0},clip]{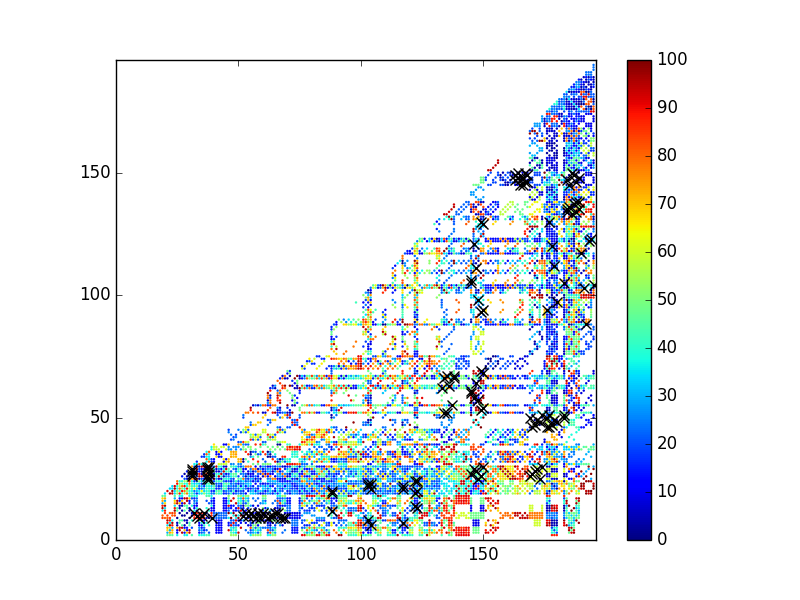}
        \caption{Normal ordering, $\chi^2<100$}
        \label{subfig:b}
\end{subfigure}
\\
\begin{subfigure}[b]{0.5\textwidth}
        \includegraphics[width=1\textwidth,trim={15mm 5mm 25mm 0},clip]{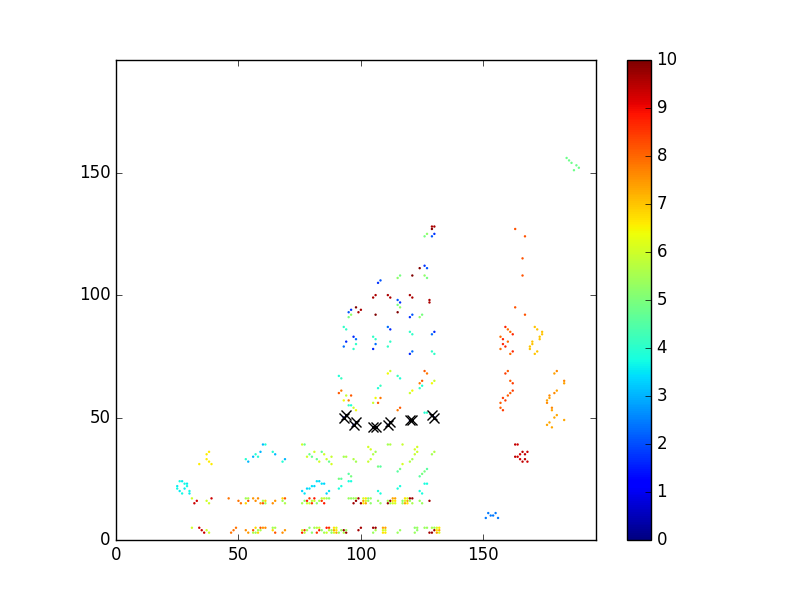}
        \caption{Inverted ordering, $\chi^2<10$}
        \label{subfig:c}
\end{subfigure}
\begin{subfigure}[b]{0.5\textwidth}
        \includegraphics[width=1\textwidth,trim={15mm 5mm 25mm 0},clip]{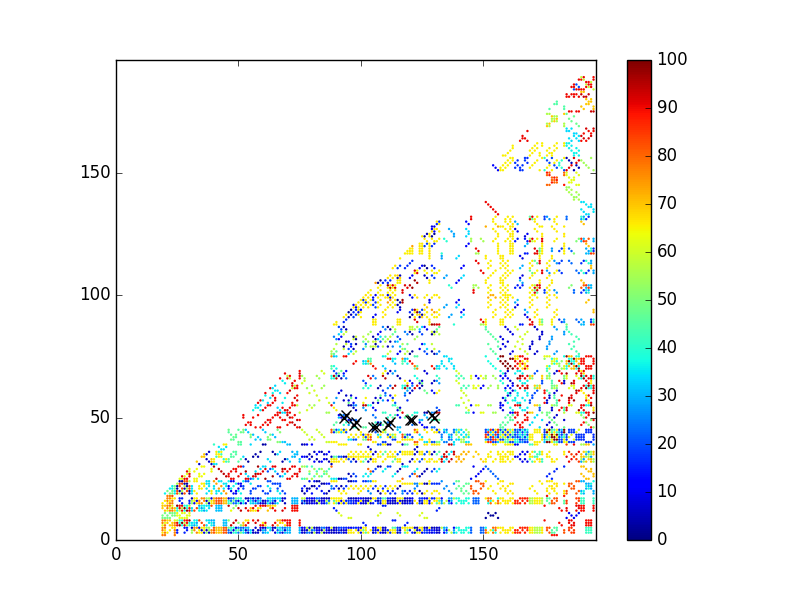}
        \caption{Inverted ordering, $\chi^2<100$}
        \label{subfig:d}
\end{subfigure}
\caption{Heat maps showing the $\chi^2$ for the $S_4$ flavour model landscape in the CSD paradigm with 2 RHNs, for each pair of EAs up to $S^{(2)}_{\text{VEV}}$, Eq. (\ref{eq:vev union}). Top row is for normal ordering and bottom row for inverted ordering. Panels on the left (right) show only pairs of EAs that have $\chi^2$ under $10$ ($100$) \label{fig:landscape}. Crosses mark the EA pairs that are viable (within the $3 \sigma$ experimental ranges).
}
\end{figure}

With all $\sigma_{x_i}$ set to the respective $3 \sigma$ range we compute the $\chi_i^2$ and their sum, $\chi^2$ and display the results for each pair of EAs in Fig.~\ref{fig:landscape} for normal ordering (top) and inverted ordering (bottom), where a Heat Map was constructed according to the $\chi^2$. In Fig.~\ref{fig:landscape}, the axes have EAs with progressively larger entries, e.g. the first EA is $(0,0,1)$ and up to EA 17 the first entry of the EA remains at zero, with EA 18 having a first entry with a small magnitude and so on. Since each EA pair leads to identical results regardless of their ordering (permuting the two EAs is equivalent to permuting $\mu_1$ with $\mu_2$), we omit the redundant results from the upper triangle of each plot in Fig.~\ref{fig:landscape}.
 
To better display the results, we show two different ranges, and additionally mark viable EA pairs (i.e. those for which our six observables fall within the $3 \sigma$ range) with black crosses. The plots on the left of Fig.~\ref{fig:landscape} $\chi^2 \leq 10$, whereas those on the right have $\chi^2 \leq 100$. We selected these ranges to enhance the presentation of the relevant results.

Fig.~\ref{fig:landscape} shows that the overwhelming majority of the EA pairs cannot achieve results which are compatible with the experimental data, returning $\chi^2 = \sum_i \chi_i^2$ orders of magnitude above the unit value. Furthermore, it is also clear that the normal ordering scenario is generally favoured over the inverted ordering scenario.
An interesting aspect that can be visualized in Fig.~\ref{fig:landscape} is the presence of some vertical and horizontal features, seen more clearly in the plots on the right. These features are indicative that some EAs are preferred by current experimental data, in the sense that those preferred EAs have (relatively) small $\chi^2$ when paired with multiple other EAs. We also see that pairs in the bottom left corner never have $\chi^2 < 100$, which is due to both EAs having the form $(0,y,z)$, which would lead to a vanishing mixing angle.

Nevertheless, a small number of cases for each scenario are able to achieve results within the $3 \sigma$ experimental range (these EA pairs are marked with black crosses). These cases are among the viable pairs (viable meaning here within the $3\sigma$ experimental range). Note that since we have allowed a row permutation in Eq.~\eqref{eq:lepton_VEV}, some of these pairs may give the same predictions, e.g., 
\begin{eqnarray} 
V_1 = c_1 \propto  \begin{pmatrix} 1 \\ -2\omega^2 \\ -2\omega \end{pmatrix}\,,~~
V_2 = (d_1^* d_4)_{\mathbf{3}_S} \propto  \begin{pmatrix} 1 \\ -2 \\ 4 \end{pmatrix}\,,
\end{eqnarray}
compared with 
\begin{eqnarray} 
V'_1 = c_2 \propto  \begin{pmatrix} 1 \\ -2\omega \\ -2\omega^2 \end{pmatrix}\,,~~
V'_2 = (d_1 d_4^*)_{\mathbf{3}_S} \propto  \begin{pmatrix} 1 \\ 4 \\ -2 \end{pmatrix} \,.
\end{eqnarray}
After extracting out this redundancy, of the $102$ pairs of EAs that are viable, we are left with 53 viable pairs, which give different predictions for neutrino masses and flavour mixing. These viable cases for normal ordering are listed in Table~\ref{tab:EAs_NH} of Appendix \ref{app:viable}.
For the inverted ordering, only $12$ pairs of EAs are viable. After extracting out the same type of redundancy, there are 6 pairs left, as shown in Table~\ref{tab:EAs_IH} of Appendix \ref{app:viable}.

We present here some examples of viable EA pairs, showing the respective contractions from VEVs (shown in \ref{eq:flavon_vevs}) that give rise to those EAs. 
\begin{eqnarray}
&&c_1 \propto \begin{pmatrix} 1 \\ -2\omega^2 \\ -2\omega \end{pmatrix}\,,~~ 
(d_1^* d_4)_{\mathbf{3}_S} \propto \begin{pmatrix} 1 \\ -2 \\ 4 \end{pmatrix} \,;  \label{eq:a+bc_example}\\
&&(a_1 c_2)_{\mathbf{3}_S} \propto \begin{pmatrix} 0 \\ 1 \\ -1 \end{pmatrix}\,,~~ 
(b_2 b_3)_{\mathbf{3}_S} \propto \begin{pmatrix} \frac{1}{3} \\ \frac{1}{3}+\frac{1}{\sqrt{3}} \\ \frac{1}{3}-\frac{1}{\sqrt{3}} \end{pmatrix}\,; \label{eq:ab+cd_example} \\
&&(a_2 c_1)_{\mathbf{3}_S} \propto \begin{pmatrix} 0 \\ 1 \\ -1 \end{pmatrix}\,,~~ 
(c_1 d_2)_{\mathbf{3}_S} \propto \begin{pmatrix} 2\omega^2 \\ 3\omega-1 \\ 2\omega \end{pmatrix}\,. 
\label{eq:ab+bc_example}
\end{eqnarray}
We note the pair in Eq.~\ref{eq:a+bc_example} is particularly interesting, since one of the EAs is one of the VEVs (in this case, $c_2$) in Eq. (\ref{eq:flavon_vevs}), i.e. part of $S^{(1)}_{\text{VEV}}$. Also noteworthy is the other EA in this pair: up to minus sign it is a permutation of the (1,4,2) direction, which we obtain here in a much simpler fashion than in the original CSD4 alignment \cite{King:2013xba}. In addition, Eqs.~\ref{eq:a+bc_example} and \ref{eq:ab+cd_example} correspond to examples with TM1 mixing \cite{Xing:2006ms, Lam:2006wm, Albright:2008rp, Albright:2010ap}.

We note also that there are no viable cases where both EAs are VEVs (part of $S^{(1)}_{\text{VEV}}$). Within the viable cases we have found three distinct topologies, represented by each of the examples in Eqs.~\ref{eq:a+bc_example}-\ref{eq:ab+bc_example}.
\begin{enumerate}
\item $\Phi_{a}+\Phi_{b} \Phi_{c}$, where one EA is a VEV $\Phi_{a}$ (part of $S^{(1)}_{\text{VEV}}$) and the other belongs to  $S^{(2)}_{\text{VEV}}$ without involving $\Phi_{a}$.
\item $\Phi_{a}\Phi_{b} + \Phi_{c} \Phi_{d}$ both EAs belong to $S^{(2)}_{\text{VEV}}$ without repeated VEVs.
\item $\Phi_{a}\Phi_{b} + \Phi_{b} \Phi_{c}$ both EAs belong to $S^{(2)}_{\text{VEV}}$ with a single repeated VEV appearing in both EAs.
\end{enumerate}
There are also models with repeated flavons such as $\Phi_{a}+\Phi_{a} \Phi_{b}$ or $\Phi_{a} \Phi_{b}+ \Phi_{b} \Phi_{a}$, but in the $S_4$ landscape none of these options are experimentally viable. Indeed of the cases with repeated flavons, $\Phi_{a}\Phi_{b} + \Phi_{b} \Phi_{c}$ leads to viable cases for normal ordering only.

\section{Examples of viable models \label{sec:models}}

We now produce some example models for each topology found in Section \ref{sec:landscape}. In the following
$\tilde{H}_\alpha$ transforms (like $H$) as an $SU(2)$ doublet, $\tilde{H}_\alpha = \epsilon_{\alpha \beta} H_\beta$.
Further, we do not specify the $S_4$ assignments here for simplicity (see Appendix \ref{app:S4} and \cite{deMedeirosVarzielas:2017hen} for more details).

\subsection{$\Phi_{a}+\Phi_{b} \Phi_{c}$}

The first topology found in Section \ref{sec:landscape} is where one of the EAs is a single flavon and the other EA is a two-flavon contraction without repeated flavons.
For this topology we write the Lagrangian with renormalisable terms
\begin{align}
\mathcal{L}_{\nu} &= \overline{\ell_L} \tilde{H} A +  \overline{A} \Phi_a N_1
+\overline{A} \Phi_b B + \overline{B} \Phi_c N_2\\
&+M_A \overline{A} A + M_B \overline{B} B
+ M_1 \overline{N_1^c} N_1 + M_2 \overline{N_2^c} N_2 + \text{h.c.} \,,
\label{eq:L_a+bc}
\end{align}
which can be implemented for example with an additional $Z_8$ with charges listed in Table \ref{A+BC_charges}.

\captionsetup{width=\textwidth}

\begin{table}[h!] \centering
    \caption{Fields and $Z_8$ charges in additive notation ($-3$ to $4$) for models with topology $\Phi_{a}+\Phi_{b} \Phi_{c}$. \label{A+BC_charges}}
    \begin{tabular}{|c|cc|cc|ccc|cc|}
      \hline\hline
      Field & $\ell_L$ & $H$ & $N_1$ & $N_2$ & $\Phi_a$ & $\Phi_b$ & $\Phi_c$ & $A$ & $B$ \\ \hline
      $Z_8$ & $-1$ & $0$ & $0$ & $4$ & $-1$ & $-3$ & $-2$ & $-1$ & $2$ \\ \hline\hline
    \end{tabular}
\end{table}

\begin{figure}
\begin{center}
\includegraphics{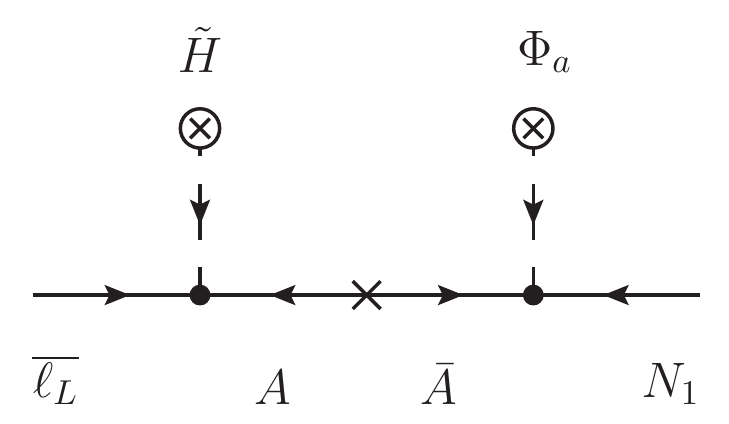}
\caption{Effective diagram for EA coming from $\Phi_a$ \label{fig:A}
}
\end{center}
\end{figure}

\begin{figure}
\begin{center}
\includegraphics{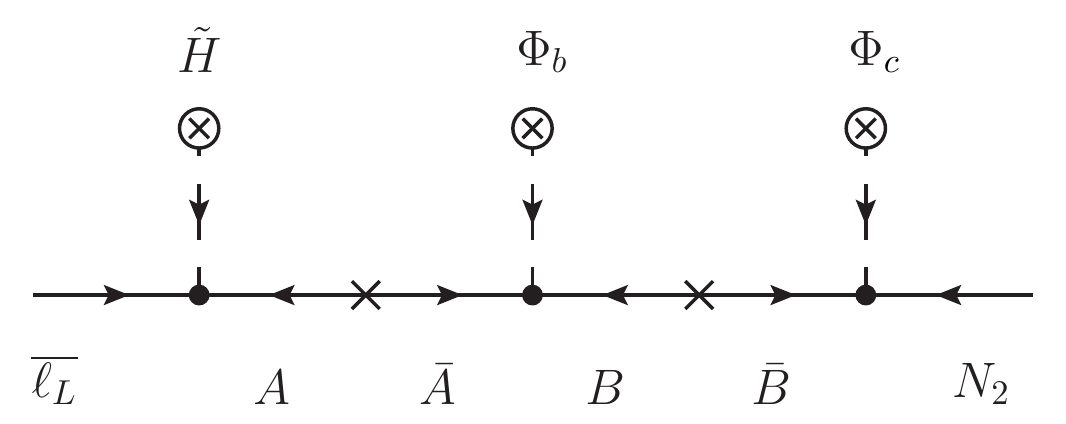}
\caption{Effective diagram for EA coming from $\Phi_b$ and $\Phi_c$ \label{fig:BC}
}
\end{center}
\end{figure}

The Lagrangian in Eq.~\ref{eq:L_a+bc} corresponds to the effective diagrams in Fig. \ref{fig:A} and Fig. \ref{fig:BC}.

One specific example of this topology is Eq.~\ref{eq:a+bc_example}, i.e. for $\Phi_{a} = c_2$, and $\Phi_{b} = d_1^*$ and $\Phi_{c} = d_4$ contracting as $(d_1^* d_4)_{\mathbf{3}_S}$ to make the EA.

\subsection{$\Phi_{a}\Phi_{b}+\Phi_{c}\Phi_{d}$}

The second topology found in Section \ref{sec:landscape} is where both EAs are two-flavon contractions without repeated flavons.
The corresponding renormalisable Lagrangian terms are
\begin{align}
\mathcal{L}_{\nu} &= \overline{\ell_L} \tilde{H} A + \overline{A} \Phi_a B + \overline{B} \Phi_b N_1
+\overline{A} \Phi_{c} C + \overline{C} \Phi_d N_2\\
&+M_A \overline{A} A + M_B \overline{B} B +M_C \overline{C} C
+ M_1 \overline{N_1^c} N_1 + M_2 \overline{N_2^c} N_2 + \text{h.c.} \,,
\label{eq:L_ab+cd}
\end{align}
which can be implemented for example with an additional $Z_{10}$ with charges listed in Table \ref{AB+CD_charges}.

\begin{table}[h!] \centering
    \caption{Fields and $Z_{8}$ charges in additive notation ($-3$ to $4$) for models with topology $\Phi_{a}\Phi_{b}+\Phi_{c}\Phi_{d}$. \label{AB+CD_charges}}
    \begin{tabular}{|c|cc|cc|cccc|ccc|}
      \hline\hline
      Field & $\ell_L$ & $H$ & $N_1$ & $N_2$ & $\Phi_a$ & $\Phi_b$ & $\Phi_c$ & $\Phi_d$ & $A$ & $B$ & $C$ \\ \hline
      $Z_{8}$ & $-1$ & $0$ & $0$ & $4$ & $4$ & $3$ & $1$ & $2$ & $-1$ & $3$ & $-2$ \\ \hline\hline
    \end{tabular}
\end{table}

\begin{figure}
\begin{center}
\includegraphics{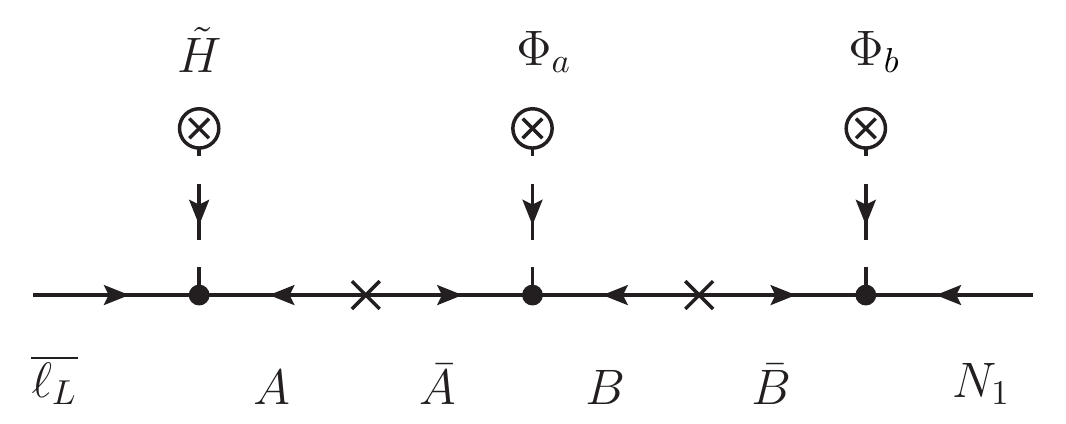}
\caption{Effective diagram for EA coming from $\Phi_a$ and $\Phi_b$. \label{fig:AB}
}
\end{center}
\end{figure}

\begin{figure}
\begin{center}
\includegraphics{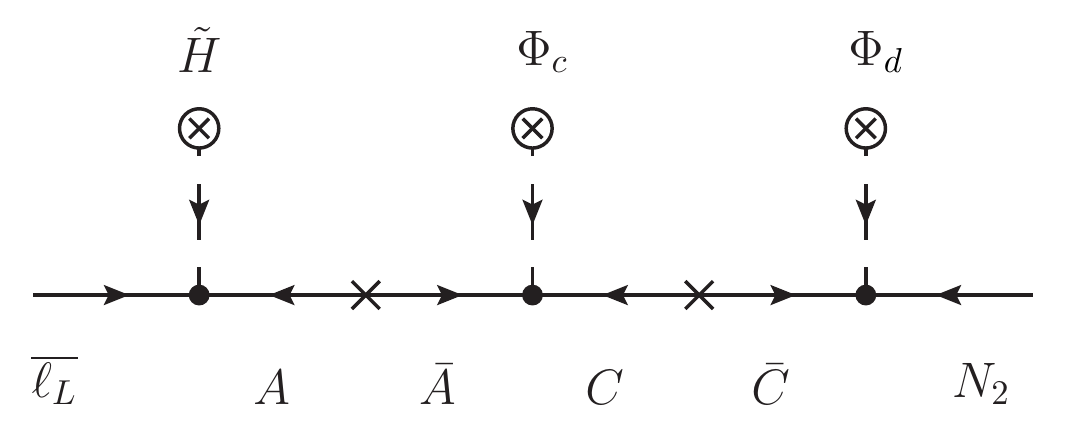}
\caption{Effective diagram for EA coming from $\Phi_c$ and $\Phi_d$. \label{fig:CD}
}
\end{center}
\end{figure}

The Lagrangian in Eq.~\ref{eq:L_ab+cd} corresponds to the effective diagrams in Fig. \ref{fig:AB} and Fig. \ref{fig:CD}.

Eq.~\ref{eq:ab+cd_example} exemplifies this topology, with
$\Phi_{a} = a_1$, $\Phi_{b} = c_2$, $\Phi_{c} = b_2$, $\Phi_{d} = b_3$
and the EAs being the contractions $(a_1 c_2)_{\mathbf{3}_S}$, $(b_2 b_3)_{\mathbf{3}_S}$.

\subsection{$\Phi_{a}\Phi_{b}+\Phi_{b}\Phi_{c}$}

The third topology found in Section \ref{sec:landscape} is where both EAs are two-flavon contractions with one repeated flavon, $\Phi_{a}\Phi_{b}+\Phi_{b}\Phi_{c}$
The renormalisable Lagrangian for this topology is
\begin{align}
\mathcal{L}_{\nu} &= \overline{\ell_L} \tilde{H} A + \overline{A} \Phi_a B + \overline{B} \Phi_b N_1
+\overline{A} \Phi_{c} C + \overline{C} \Phi_d N_2\\
&+M_A \overline{A} A + M_B \overline{B} B +M_C \overline{C} C
+ M_1 \overline{N_1^c} N_1 + M_2 \overline{N_2^c} N_2 + \text{h.c.}\,,
\label{eq:L_ab+cd}
\end{align}
which can be implemented for example with an additional $Z_6$ with charges listed in Table \ref{AB+BC_charges}.

\begin{table}[h!] \centering
    \caption{Fields and $Z_6$ charges in additive notation ($-2$ to $3$) for models with topology $\Phi_{a}\Phi_{b}+\Phi_{b}\Phi_{c}$. \label{AB+BC_charges}}
    \begin{tabular}{|c|cc|cc|ccc|ccc|}
      \hline\hline
      Field & $\ell_L$ & $H$ & $N_1$ & $N_2$ & $\Phi_a$ & $\Phi_b$ & $\Phi_c$ & $A$ & $B$ & $C$ \\ \hline
      $Z_6$ & $-1$ & $0$ & $0$ & $3$ & $2$ & $3$ & $1$ & $1$ & $3$ & $2$ \\ \hline\hline
    \end{tabular}
\end{table}

The Lagrangian in Eq.~\ref{eq:L_ab+cd} corresponds to effective diagrams in Fig. \ref{fig:AB} and Fig. \ref{fig:BC}. The same diagrams already appeared in the other topologies, the fields take different charges according to the respective topology.

Eq.~\ref{eq:ab+bc_example} exemplifies this topology, with EAs contracted 
$\Phi_{a} = a_1$, $\Phi_{b} = d_2^*$, $\Phi_{c} = d_2$ and the EAs being $(a_1 d_2^*)_{\mathbf{3}_S}$, $(d_2^* d_1)_{\mathbf{3}_S}$.

\section{Conclusions \label{sec:Conclusions}}

We have explored the concept of Effective Alignments (EAs) introduced in \cite{deMedeirosVarzielas:2017hen}, that is of directions in flavour space that can be constructed from contractions of Vacuum Expectation Values (VEVs).

As in \cite{deMedeirosVarzielas:2017hen}, we have considered an $S_4$ flavour symmetry as an example. Here, we considered the 21 VEVs of $S_4$, from which a total of 195 distinct EAs are obtained with contractions of up to two flavons.

Assuming the charged lepton Yukawa couplings to be diagonal in a convenient basis (which can be easily obtained in $S_4$ \cite{deMedeirosVarzielas:2017hen}), we then take the paradigm of Constrained Sequential Dominance leptonic flavour models with two right-handed neutrinos, and consider the set of all such models that can be built with any pair of the 195 EAs of $S_4$.

We test all these models to find which pairs of EAs are phenomenologically viable (meaning all observables within the respective $3 \sigma$ ranges). In total we found 102 cases for normal ordering, corresponding to 53 mixing patterns (listed in Table~\ref{tab:EAs_NH}) and 12 cases for inverted ordering, corresponding to 6 mixing patterns (listed Table~\ref{tab:EAs_IH}). Analysing our results, we found no viable models where both EAs are VEVs, but there are cases in a topology with one EA being a VEV (the other being a contraction of two VEVs). Viable cases where both EAs are contractions of two VEVs exist in a topology without repeated VEVs and also in a topology where a given VEV appears in both EAs.
Within this framework, we construct example models for each type of viable model, showing the Lagrangian, field content and charge assignments, and effective diagrams for each topology.

This type of study can be further generalized, for example considering models with more flavon contractions or even models based on other flavour symmetries.

\section*{Acknowledgements}

IdMV thanks the University of Southampton for hospitality during work on this project.

IdMV acknowledges funding from the Funda\c{c}\~{a}o para a Ci\^{e}ncia e a Tecnologia (FCT) through
the contract IF/00816/2015. IdMV acknowledges partial support by the HARMONIA project under contract
UMO-2015/18/M/ST2/00518 (2016-2019), and by FCT through projects CFTP-FCT Unit 777 (UID/FIS/00777/2013), CERN/FIS-PAR/0004/2017 and PTDC/FIS-PAR/29436/2017 which are partially funded through POCTI (FEDER), COMPETE, QREN and EU. YLZ acknowledges the STFC Consolidated Grant ST/L000296/1 and the European Union's Horizon 2020 Research and Innovation programme under Marie Sk\l{}odowska-Curie grant agreements Elusives ITN No.\ 674896 and InvisiblesPlus RISE No.\ 690575.

\appendix
\section{Group theory of $S_4$ \label{app:S4}} 

$S_4$ is the permutation group of four objects, see e.g.,\ \cite{Escobar:2008vc,Li:2014eia,Ding:2013eca}. It has 24 distinct elements,
which are generated by three generators $S$, $T$ and $U$ satisfying
the relations
\begin{equation} \label{eq:generators} 
T^3=S^2=U^2=(ST)^3=(SU)^2=(TU)^2=1\,.
\end{equation}
The Kronecker products between different irreducible representations can be easily obtained:
\begin{eqnarray}
&\mathbf{1^{\prime}}\otimes\mathbf{1^{\prime}}=\mathbf{1}, ~~\mathbf{1^{\prime}}\otimes\mathbf{2}=\mathbf{2}, ~~\mathbf{1^{\prime}}\otimes\mathbf{3}=\mathbf{3^{\prime}}, ~~
\mathbf{1^{\prime}}\otimes\mathbf{3^{\prime}}=\mathbf{3},~~\mathbf{2}\otimes\mathbf{2}=\mathbf{1}\oplus\mathbf{1}^{\prime}\oplus\mathbf{2},\nonumber\\
&
\mathbf{2}\otimes\mathbf{3}=\mathbf{2}\otimes\mathbf{3^{\prime}}=\mathbf{3}\oplus\mathbf{3}^{\prime},~~
\mathbf{3}\otimes\mathbf{3}=\mathbf{3^{\prime}}\otimes\mathbf{3^{\prime}}=\mathbf{1}\oplus \mathbf{2}\oplus\mathbf{3}\oplus\mathbf{3^{\prime}},~~
\mathbf{3}\otimes\mathbf{3^{\prime}}=\mathbf{1^{\prime}}\oplus \mathbf{2}\oplus\mathbf{3}\oplus\mathbf{3^{\prime}}
\end{eqnarray}
%

\begin{table}[h!]
\centering
\caption{\label{tab:rep_matrix} The representation matrices for the $S_4$ generators $T$, $S$ and $U$, where $\omega=e^{2\pi i/3}$.}
\begin{tabular}{|c|ccc|}
\hline\hline
   & $T$ & $S$ & $U$  \\\hline
$\mathbf{1}$ & 1 & 1 & 1 \\
$\mathbf{1^{\prime}}$ & 1 & 1 & -1 \\
$\mathbf{2}$ & 
$\left(
\begin{array}{cc}
 \omega  & 0 \\
 0 & \omega ^2 \\
\end{array}
\right)$ & 
$\left(
\begin{array}{cc}
 1 & 0 \\
 0 & 1 \\
\end{array}
\right)$ & 
$\left(
\begin{array}{cc}
 0 & 1 \\
 1 & 0 \\
\end{array}
\right)$ \\

$\mathbf{3}$ &  $\left(
\begin{array}{ccc}
 1 & 0 & 0 \\
 0 & \omega ^2 & 0 \\
 0 & 0 & \omega  \\
\end{array}
\right)$ &
$\frac{1}{3} \left(
\begin{array}{ccc}
 -1 & 2 & 2 \\
 2 & -1 & 2 \\
 2 & 2 & -1 \\
\end{array}
\right)$ &
$\left(
\begin{array}{ccc}
 1 & 0 & 0 \\
 0 & 0 & 1 \\
 0 & 1 & 0 \\
\end{array}
\right)$ \\

$\mathbf{3^{\prime}}$ &  $\left(
\begin{array}{ccc}
 1 & 0 & 0 \\
 0 & \omega ^2 & 0 \\
 0 & 0 & \omega  \\
\end{array}
\right)$ &
$\frac{1}{3} \left(
\begin{array}{ccc}
 -1 & 2 & 2 \\
 2 & -1 & 2 \\
 2 & 2 & -1 \\
\end{array}
\right)$ &
$-\left(
\begin{array}{ccc}
 1 & 0 & 0 \\
 0 & 0 & 1 \\
 0 & 1 & 0 \\
\end{array}
\right)$ \\ \hline\hline
\end{tabular}
\end{table}
 
The generators $T$, $S$, and $U$ in different irreducible representations are listed in Table \ref{tab:rep_matrix}.  
This basis is widely used in the literature since the charged lepton mass matrix invariant under $T$ is diagonal in this basis. The products of two 3 dimensional irreducible representations $a=(a_1, a_2, a_3)^T$ and $b=(b_1, b_2, b_3)^T$ can be expressed as
\begin{eqnarray}
(ab)_\mathbf{1_i} &=& a_1b_1 + a_2b_3 + a_3b_2 \,,\nonumber\\
(ab)_\mathbf{2} &=& (a_3b_3 + a_1b_2 + a_2b_1,~ a_2b_2 + a_1b_3 + a_3b_1)^T \,,\nonumber\\
(ab)_{\mathbf{3_i}} &=& \frac{1}{\sqrt{6}} (2a_1b_1-a_2b_3-a_3b_2, 2a_3b_3-a_1b_2-a_2b_1, 2a_2b_2-a_3b_1-a_1b_3)^T \,,\nonumber\\
(ab)_{\mathbf{3_j}} &=& \frac{1}{\sqrt{2}} (a_2b_3-a_3b_2, a_1b_2-a_2b_1, a_3b_1-a_1b_3)^T \,,
\label{eq:CG2}
\end{eqnarray}
where 
\begin{eqnarray}
&&\mathbf{1_i}=\mathbf{1}\,, ~\; \mathbf{3_i}=\mathbf{3}\,, ~\; \mathbf{3_j}=\mathbf{3'}\,~\; \text{for} ~\; a\sim b \sim \mathbf{3}\,,~ \mathbf{3^\prime} \,, \nonumber\\
&&\mathbf{1_i}=\mathbf{1'}\,,~  \mathbf{3_i}=\mathbf{3'}\,,~ \mathbf{3_j}=\mathbf{3}\,~\;\; \text{for} ~\; a\sim  \mathbf{3}\,,~ b \sim \mathbf{3'}\,.
\end{eqnarray} 
The products of two doublets $a=(a_1, a_2)^T$ and $b=(b_1, b_2)^T$ are ruduced to
\begin{eqnarray}
(ab)_\mathbf{1} &=& a_1b_2 + a_2b_1 \,,\quad
(ab)_\mathbf{1^\prime} = a_1b_2 - a_2b_1 \,,\quad
(ab)_{\mathbf{2}} = (a_2b_2, a_1b_1)^T \,.
\label{eq:CG_doublets}
\end{eqnarray} 

The Kronecker products of multiplets of $S_4$ require the following properties: if the trilinear combination of three multiplets $a\sim \mathbf{r}$, $b\sim \mathbf{r'}$ and $c\sim \mathbf{r''}$ is an invariance of $S_4$, e.g., $\big((ab)_\mathbf{r''}c\big)_\mathbf{1}$, then 
\begin{eqnarray}
\big((ab)_{\mathbf{r}_c}c\big)_\mathbf{1}\,=\big((bc)_{\mathbf{r}_a}a\big)_\mathbf{1}\;=\big((ca)_{\mathbf{r}_b}b\big)_\mathbf{1} \,
=\big(a(bc)_{\mathbf{r}_a}\big)_\mathbf{1}\;=\big(b(ca)_{\mathbf{r}_b}\big)_\mathbf{1}\, =\big(c(ab)_{\mathbf{r}_c}\big)_\mathbf{1}\,\,, \nonumber\\
\big((ab)_{\mathbf{r}'_c}c\big)_\mathbf{1'}=\big((bc)_{\mathbf{r}'_a}a\big)_\mathbf{1'}=\big((ca)_{\mathbf{r}'_b}b\big)_\mathbf{1'} 
=\big(a(bc)_{\mathbf{r}'_a}\big)_\mathbf{1'}=\big(b(ca)_{\mathbf{r}'_b}\big)_\mathbf{1'} =\big(c(ab)_{\mathbf{r}'_c}\big)_\mathbf{1'}\,.
\end{eqnarray}
holds, where $\mathbf{r}'=\mathbf{1'},\mathbf{1},\mathbf{2},\mathbf{3'},\mathbf{3}$ for $\mathbf{r}=\mathbf{1},\mathbf{1'},\mathbf{2},\mathbf{3},\mathbf{3'}$, respectively \cite{deMedeirosVarzielas:2017hen}. This property is used for flavon contractions in section~\ref{sec:models}. 

\section{Viable pairs of EAs \label{app:viable}}

In this appendix, we list analytical expressions of EAs which are compatible with oscillation data in $3\sigma$ ranges. Results in the normal mass ordering are shown in Table~\ref{tab:EAs_NH}, and those in the inverted mass ordering are shown in Table~\ref{tab:EAs_IH}. All vectors are normalised to 1. The phase difference between $e^{i\alpha_1}V_1$ and $V_1$, and that between $e^{i\alpha_2} V_2$ and $V_2$ (for arbitrary $\alpha_1$ and $\alpha_2$) can be absorbed by the redefinition of the parameters $\mu_1$ and $\mu_2$, respectively and thus $e^{i\alpha_1}V_1$ and $e^{i\alpha_2} V_2$ do not need to be listed independently from $V_1$ and $V_2$. 

\begin{longtable}[t!]{p{9.2cm}p{7.5cm}}
\caption{The two EAs compatible with data in $3\sigma$ ranges in the normal mass ordering. Each EA as a flavon VEV or a contraction of of two flavon VEVs (not unique) is also shown.} \label{tab:EAs_NH}\\
\hline\hline

$V_1$ & $V_2$ \\\hline\\[-4mm] \endhead

\hline \multicolumn{2}{r}{{Continued on next page}}
\endfoot
\hline\hline
\endlastfoot

$ \left(0,\frac{1}{\sqrt{5}},\frac{2}{\sqrt{5}}\right)^T = (c_4 d_1)_{\mathbf{3}_A}$ & $\left(\frac{1}{\sqrt{3}},-\frac{\left(-3+\sqrt{3}\right) \omega^2}{6} ,-\frac{\left(3+\sqrt{3}\right) \omega}{6} \right)^T = b_2$ \\
$ \left(0,\frac{\omega^2}{\sqrt{5}},\frac{2\omega}{\sqrt{5}}\right)^T = (c_4 d_2)_{\mathbf{3}_A}$ & $\left(\frac{1}{\sqrt{3}},-\frac{\left(-3+\sqrt{3}\right) \omega}{6} ,-\frac{\left(3+\sqrt{3}\right) \omega^2}{6} \right)^T=b_3$ \\
$ \left(0,\frac{-2\omega^2}{\sqrt{5}},\frac{-\omega}{\sqrt{5}}\right)^T = (c_4 d_2^*)_{\mathbf{3}_A}$ & $\left(\frac{1}{\sqrt{3}},\frac{-3-\sqrt{3}}{6} ,\frac{3-\sqrt{3}}{6} \right)^T=b_1^*$ \\
$ \left(0,\frac{-2\omega}{\sqrt{5}},\frac{-\omega^2}{\sqrt{5}}\right)^T = (c_4 d_3^*)_{\mathbf{3}_A}$ & $\left(\frac{1}{\sqrt{3}},-\frac{\left(3+\sqrt{3}\right) \omega^2}{6} ,-\frac{\left(-3+\sqrt{3}\right) \omega}{6} \right)^T =b_2^*$ \\

$ \left(0,-\frac{1}{\sqrt{2}},\frac{1}{\sqrt{2}}\right)^T = (a_1 c_2)_{\mathbf{3}_S}$ & $\left(\frac{1}{3},\frac{1}{3}+\frac{1}{\sqrt{3}},\frac{1}{3}-\frac{1}{\sqrt{3}}\right)^T = (b_2 b_3)_{\mathbf{3}_S}$ \\
$ \left(0,-\frac{1}{\sqrt{2}},\frac{1}{\sqrt{2}}\right)^T = (a_1 c_2)_{\mathbf{3}_S}$ & $\left(\frac{2\omega^2}{\sqrt{21}},\frac{2}{\sqrt{21}},-\frac{3+4 \omega}{\sqrt{21}}\right)^T = (c_2 d_1^*)_{\mathbf{3}_S}$ \\
$ \left(-\frac{1}{\sqrt{6}},-\frac{1}{\sqrt{6}},\sqrt{\frac{2}{3}}\right)^T = (a_1 d_1)_{\mathbf{3}_S}$ & $\left(\sqrt{\frac{2}{3}},\frac{\left(-3+\sqrt{3}\right) \omega^2}{6 \sqrt{2}},\frac{\left(3+\sqrt{3}\right) \omega}{6 \sqrt{2}}\right)^T = (b_2 c_4)_{\mathbf{3}_S}$ \\
$ \left(-\frac{1}{\sqrt{6}},-\frac{1}{\sqrt{6}},\sqrt{\frac{2}{3}}\right)^T = (a_1 d_1)_{\mathbf{3}_S}$ & $\left(\sqrt{\frac{2}{3}},\frac{\left(-3+\sqrt{3}\right) \omega}{6 \sqrt{2}},\frac{\left(3+\sqrt{3}\right) \omega^2}{6 \sqrt{2}}\right)^T = (b_3 c_4)_{\mathbf{3}_S}$ \\

$ \left(0,\frac{1}{\sqrt{2}},-\frac{1}{\sqrt{2}}\right)^T = (a_2 c_1)_{\mathbf{3}_S}$ & $\left(\frac{1}{3},\frac{\left(1+\sqrt{3}\right) \omega^2}{3} ,-\frac{\left(-1+\sqrt{3}\right) \omega}{3} \right)^T = (b_1 b_3)_{\mathbf{3}_S}$ \\
$ \left(0,\frac{1}{\sqrt{2}},-\frac{1}{\sqrt{2}}\right)^T = (a_2 c_1)_{\mathbf{3}_S}$ & $\left(\frac{1}{3},-\frac{\left(-1+\sqrt{3}\right) \omega^2}{3} ,\frac{\left(1+\sqrt{3}\right) \omega}{3} \right)^T = (b_1^* b_3^*)_{\mathbf{3}_S}$ \\
$ \left(0,\frac{1}{\sqrt{2}},-\frac{1}{\sqrt{2}}\right)^T = (a_2 c_1)_{\mathbf{3}_S}$ & $\left(\frac{2 \omega^2}{\sqrt{21}},\frac{3 \omega -1}{\sqrt{21}},\frac{2 \omega}{\sqrt{21}}\right)^T = (c_1 d_2)_{\mathbf{3}_S}$ \\
$ \left(0,\frac{1}{\sqrt{2}},-\frac{1}{\sqrt{2}}\right)^T = (a_2 c_1)_{\mathbf{3}_S}$ & $\left(\frac{2 \omega^2}{\sqrt{21}},\frac{2 \omega^2}{\sqrt{21}},\frac{3-\omega^2}{\sqrt{21}}\right)^T = (c_3 d_2^*)_{\mathbf{3}_S}$ \\
$ \left(\frac{1}{\sqrt{2}},0,-\frac{1}{\sqrt{2}}\right)^T = (a_2 d_1^*)_{\mathbf{3}_S}$ & $\left(\frac{3-\omega^2}{\sqrt{21}},\frac{2 \omega^2}{\sqrt{21}},\frac{2 \omega^2}{\sqrt{21}}\right)^T = (d_2 d_3^*)_{\mathbf{3}_S}$ \\
$ \left(-\frac{1}{\sqrt{6}},-\frac{\omega^2}{\sqrt{6}},\sqrt{\frac{2}{3}} \omega\right)^T = (a_2 d_2)_{\mathbf{3}_S}$ & $\left(\sqrt{\frac{2}{3}},\frac{-3+\sqrt{3}}{6 \sqrt{2}},\frac{3+\sqrt{3}}{6 \sqrt{2}}\right)^T = (b_1 c_4)_{\mathbf{3}_S}$ \\
$ \left(-\frac{1}{\sqrt{6}},-\frac{\omega^2}{\sqrt{6}},\sqrt{\frac{2}{3}} \omega\right)^T = (a_2 d_2)_{\mathbf{3}_S}$ & $\left(\sqrt{\frac{2}{3}},\frac{(-3+\sqrt{3})\omega}{6 \sqrt{2}},\frac{(3+\sqrt{3})\omega^2}{6 \sqrt{2}}\right)^T = (b_3 c_4)_{\mathbf{3}_S}$ \\
$ \left(-\frac{1}{\sqrt{6}},\sqrt{\frac{2}{3}}\omega^2,-\frac{\omega}{\sqrt{6}}\right)^T = (a_2 d_2^*)_{\mathbf{3}_S}$ & $\left(\sqrt{\frac{2}{3}},\frac{3+\sqrt{3}}{6 \sqrt{2}},\frac{-3+\sqrt{3}}{6 \sqrt{2}}\right)^T = (b_1^* c_4)_{\mathbf{3}_S}$ \\
$ \left(-\frac{1}{\sqrt{6}},\sqrt{\frac{2}{3}}\omega^2,-\frac{\omega}{\sqrt{6}}\right)^T = (a_2 d_2^*)_{\mathbf{3}_S}$ & $\left(\sqrt{\frac{2}{3}},\frac{(3+\sqrt{3})\omega}{6 \sqrt{2}},\frac{(-3+\sqrt{3})\omega^2}{6 \sqrt{2}}\right)^T = (b_3^* c_4)_{\mathbf{3}_S}$ \\

$ \left(0,\frac{1}{\sqrt{2}},-\frac{1}{\sqrt{2}}\right)^T = (a_3 c_1)_{\mathbf{3}_S}$ & $\left(\frac{1}{3},\frac{\left(1+\sqrt{3}\right) \omega}{3} ,-\frac{\left(-1+\sqrt{3}\right) \omega^2}{3} \right)^T = (b_1 b_2)_{\mathbf{3}_S}$ \\
$ \left(0,\frac{1}{\sqrt{2}},-\frac{1}{\sqrt{2}}\right)^T = (a_3 c_1)_{\mathbf{3}_S}$ & $\left(\frac{1}{3},-\frac{\left(-1+\sqrt{3}\right) \omega}{3} ,\frac{\left(1+\sqrt{3}\right) \omega^2}{3} \right)^T = (b_1^* b_2^*)_{\mathbf{3}_S}$ \\
$ \left(0,\frac{1}{\sqrt{2}},-\frac{1}{\sqrt{2}}\right)^T = (a_3 c_1)_{\mathbf{3}_S}$ & $\left(\frac{2 \omega^2}{\sqrt{21}},\frac{2 \omega}{\sqrt{21}},\frac{3\omega^2-1}{\sqrt{21}}\right)^T = (c_1 d_3^*)_{\mathbf{3}_S}$ \\

$ \left(\frac{1-i}{\sqrt{6}},-\frac{\sqrt{3} (1+2 i)+3}{6 \sqrt{2}},\frac{\left(3 \sqrt{2}-(1+2 i) \sqrt{6}\right)}{12} \right)^T = (b_1 c_3)_{\mathbf{3}_S}$ & $\left(\frac{1}{\sqrt{2}},\frac{-1}{\sqrt{2}},0\right)^T = (a_1 d_2)_{\mathbf{3}_S}$ \\
$ \left(\frac{\left(-3+\sqrt{3}\right)}{3 \sqrt{2 \left(4+\sqrt{3}\right)}},-\frac{2\left(3+\sqrt{3}\right)}{3\sqrt{2(4+\sqrt{3})}},-\sqrt{\frac{2}{3 \left(4+\sqrt{3}\right)}} \right)^T = (b_1 d_4^*)_{\mathbf{3}_S}$ & $\left(\frac{1}{\sqrt{3}},-\frac{(3+\sqrt{3})\omega}{6},-\frac{(-3+\sqrt{3})\omega^2}{6}\right)^T = b_3^*$ \\

$ \left(\frac{1-i}{\sqrt{6}},\frac{(-3+3 i)+\sqrt{3} (1+5 i)}{12 \sqrt{2}},\frac{\sqrt{3} (1+5 i)+(3-3 i)}{12 \sqrt{2}}\right)^T = (b_2 c_1)_{\mathbf{3}_S}$ & $\left(\frac{1}{\sqrt{2}},-\frac{\omega^2}{\sqrt{2}},0\right)^T = (a_2 d_1)_{\mathbf{3}_S}$ \\
$ \left(\frac{\left(-3+\sqrt{3}\right) \omega^2}{3 \sqrt{2 \left(4+\sqrt{3}\right)}},-\frac{2\left(3+\sqrt{3}\right)\omega}{3\sqrt{2(4+\sqrt{3})}},-\sqrt{\frac{2}{3 \left(4+\sqrt{3}\right)}} \right)^T = (b_2 d_4^*)_{\mathbf{3}_S}$ & $\left(\frac{1}{\sqrt{3}},\frac{\left(-3-\sqrt{3}\right)}{6} ,\frac{\left(3-\sqrt{3}\right)}{6} \right)^T = b_1^*$ \\

$ \left(\frac{1-i}{\sqrt{6}},\frac{\sqrt{3} (1-i)+(-9+3 i)}{12 \sqrt{2}},\frac{(1-i) \left(\sqrt{3}+(6+3 i)\right)}{12 \sqrt{2}}\right)^T = (b_2^* c_3)_{\mathbf{3}_S}$ & $\left(\frac{1}{\sqrt{2}},0,-\frac{\omega}{\sqrt{2}}\right)^T = (a_2 d_1^*)_{\mathbf{3}_S}$ \\
$ \left(\frac{\left(-3+\sqrt{3}\right) \omega}{3 \sqrt{2 \left(4+\sqrt{3}\right)}},-\sqrt{\frac{2}{3 \left(4+\sqrt{3}\right)}},-\frac{2\left(3+\sqrt{3}\right)\omega^2}{3\sqrt{2(4+\sqrt{3})}} \right)^T = (b_2^* d_4)_{\mathbf{3}_S}$ & $\left(\frac{1}{3},-\frac{(-3+\sqrt{3})\omega}{6},-\frac{(3+\sqrt{3})\omega^2}{6}\right)^T = b_3$ \\

$ \left(\frac{2 \omega^2}{\sqrt{21}},\frac{3\omega^2-1}{\sqrt{21}},\frac{2 \omega}{\sqrt{21}}\right)^T = (c_1 d_2)_{\mathbf{3}_S}$ & $\left(\frac{1}{\sqrt{2}},-\frac{1}{\sqrt{2}},0\right)^T = (a_1 d_2)_{\mathbf{3}_S}$ \\
$ \left(\frac{2}{\sqrt{21}},-\frac{1}{\sqrt{21}},-\frac{4}{\sqrt{21}}\right)^T = (c_1 d_4)_{\mathbf{3}_S}$ & $\left(-\frac{2}{3},\frac{\omega^2}{3},-\frac{2}{3} \omega\right)^T = d_2$ \\
$ \left(\frac{2}{\sqrt{21}},-\frac{1}{\sqrt{21}},-\frac{4}{\sqrt{21}}\right)^T = (c_1 d_4)_{\mathbf{3}_S}$ & $\left(-\frac{2}{3},\frac{\omega}{3},-\frac{2}{3} \omega^2\right)^T = d_3$ \\

$ \left(\frac{2 \omega^2}{\sqrt{21}},\frac{2}{\sqrt{21}},-\frac{3+4 \omega^2}{\sqrt{21}}\right)^T = (c_2 d_1^*)_{\mathbf{3}_S}$ & $\left(\frac{1}{\sqrt{2}},0,-\frac{\omega}{\sqrt{2}}\right)^T = (a_2 d_1^*)_{\mathbf{3}_S}$ \\
$ \left(\frac{2 \omega^2}{\sqrt{21}},\frac{3-\omega^2}{\sqrt{21}},\frac{2 \omega^2}{\sqrt{21}}\right)^T = (c_2 d_3)_{\mathbf{3}_S}$ & $\left(\frac{1}{\sqrt{2}},-\frac{\omega^2}{\sqrt{2}},0\right)^T = (a_2 d_1)_{\mathbf{3}_S}$ \\

$ \left(\frac{2 \omega}{\sqrt{21}},-\frac{1}{\sqrt{21}},-\frac{4 \omega^2}{\sqrt{21}}\right)^T = (c_2 d_4)_{\mathbf{3}_S}$ & $\left(-\frac{2}{3},\frac{1}{3},-\frac{2}{3}\right)^T = d_1^*$ \\

$ \left(\frac{2 \omega}{\sqrt{21}},-\frac{1}{\sqrt{21}},-\frac{4 \omega^2}{\sqrt{21}}\right)^T = (c_2 d_4)_{\mathbf{3}_S}$ & $\left(-\frac{2}{3},\frac{\omega}{3},-\frac{2}{3} \omega^2\right)^T = d_3$ \\

$ \left(\frac{2 \omega^2}{\sqrt{21}},-\frac{4 \omega}{\sqrt{21}},-\frac{1}{\sqrt{21}}\right)^T = (c_2 d_4^*)_{\mathbf{3}_S}$ & $\left(-\frac{2}{3},\frac{1}{3},-\frac{2}{3}\right)^T = d_1$ \\

$ \left(\frac{2 \omega^2}{\sqrt{21}},-\frac{4 \omega}{\sqrt{21}},-\frac{1}{\sqrt{21}}\right)^T = (c_2 d_4^*)_{\mathbf{3}_S}$ & $\left(-\frac{2}{3},-\frac{2\omega}{3},\frac{\omega^2}{3}\right)^T = d_3^*$ \\

$ \left(-\frac{1}{\sqrt{21}},-\frac{4}{\sqrt{21}},\frac{2}{\sqrt{21}}\right)^T = (d_1 d_4^*)_{\mathbf{3}_S}$ & $\left(\frac{1}{3},-\frac{2}{3} \omega^2,-\frac{2}{3} \omega\right)^T = c_2$ \\
$ \left(-\frac{1}{\sqrt{21}},-\frac{4}{\sqrt{21}},\frac{2}{\sqrt{21}}\right)^T = (d_1 d_4^*)_{\mathbf{3}_S}$ & $\left(\frac{1}{3},-\frac{2}{3} \omega,-\frac{2}{3} \omega^2\right)^T = c_3$ \\
$ \left(-\frac{1}{\sqrt{21}},-\frac{4}{\sqrt{21}},\frac{2}{\sqrt{21}}\right)^T = (d_1 d_4^*)_{\mathbf{3}_S}$ & $\left(-\frac{1+\sqrt{3}}{\sqrt{2 \left(4+\sqrt{3}\right)}},-\sqrt{\frac{2}{4+\sqrt{3}}},0\right)^T = (b_1^* d_1)_{\mathbf{3}_S}$ \\

$ \left(-\frac{\omega^2}{\sqrt{21}},-\frac{4 \omega}{\sqrt{21}},\frac{2}{\sqrt{21}}\right)^T = (d_2 d_4^*)_{\mathbf{3}_S}$ & $\left(\frac{1}{3},-\frac{2}{3} \omega^2,-\frac{2}{3} \omega\right)^T = c_1$ \\
$ \left(-\frac{\omega^2}{\sqrt{21}},-\frac{4 \omega}{\sqrt{21}},\frac{2}{\sqrt{21}}\right)^T = (d_2 d_4^*)_{\mathbf{3}_S}$ & $\left(\frac{1}{3},-\frac{2}{3} \omega,-\frac{2}{3} \omega^2\right)^T = c_3$ \\
$ \left(-\frac{\omega^2}{\sqrt{21}},-\frac{4 \omega}{\sqrt{21}},\frac{2}{\sqrt{21}}\right)^T = (d_2 d_4^*)_{\mathbf{3}_S}$ & $\left(-\frac{1+\sqrt{3}}{\sqrt{2 \left(4+\sqrt{3}\right)}},-\sqrt{\frac{2}{4+\sqrt{3}}} \omega^2,0\right)^T = (b_2^* d_2)_{\mathbf{3}_S}$ \\

$ \left(-\frac{\omega}{\sqrt{21}},\frac{2}{\sqrt{21}},-\frac{4 \omega^2}{\sqrt{21}}\right)^T = (d_2^* d_4)_{\mathbf{3}_S}$ & $\left(\frac{1}{3},-\frac{2}{3} \omega^2,-\frac{2}{3} \omega\right)^T = c_1$ \\
$ \left(-\frac{\omega}{\sqrt{21}},\frac{2}{\sqrt{21}},-\frac{4 \omega^2}{\sqrt{21}}\right)^T = (d_2^* d_4)_{\mathbf{3}_S}$ & $\left(\frac{1}{3},-\frac{2}{3} \omega,-\frac{2}{3} \omega^2\right)^T = c_3$ \\
$ \left(-\frac{\omega}{\sqrt{21}},\frac{2}{\sqrt{21}},-\frac{4 \omega^2}{\sqrt{21}}\right)^T = (d_2^* d_4)_{\mathbf{3}_S}$ & $\left(-\frac{1+\sqrt{3}}{\sqrt{2 \left(4+\sqrt{3}\right)}},0,-\sqrt{\frac{2}{4+\sqrt{3}}} \omega\right)^T = (b_2 d_2^*)_{\mathbf{3}_S}$ \\

$ \left(\frac{1}{\sqrt{3}},\frac{\left(3-\sqrt{3}\right)}{6} ,\frac{\left(-3-\sqrt{3}\right)}{6} \right)^T = b_1$ & $\left(\frac{2\omega}{\sqrt{5}},\frac{1}{\sqrt{5}},0\right)^T = (c_2 d_4)_{\mathbf{3}_A}$ \\
$ \left(\frac{1}{\sqrt{3}},-\frac{\left(-3+\sqrt{3}\right) \omega^2}{6} ,-\frac{\left(3+\sqrt{3}\right) \omega}{6} \right)^T = b_2$ & $\left(\frac{2\omega^2}{\sqrt{5}},\frac{1}{\sqrt{5}},0\right)^T = (c_3 d_4)_{\mathbf{3}_A}$ \\
$ \left(\frac{1}{\sqrt{3}},-\frac{\left(-3+\sqrt{3}\right) \omega}{6} ,-\frac{\left(3+\sqrt{3}\right) \omega^2}{6} \right)^T = b_2^*$ & $\left(\frac{2}{\sqrt{5}},0,\frac{1}{\sqrt{5}}\right)^T = (c_1 d_4^*)_{\mathbf{3}_A}$ \\

$ \left(-\frac{2}{3},\frac{1}{3},-\frac{2}{3}\right)^T = d_1$ & $\left(-\frac{4}{\sqrt{21}},-\frac{\omega^2}{\sqrt{21}},\frac{2\omega}{\sqrt{21}}\right)^T = (c_4 d_2)_{\mathbf{3}_S}$ \\
$ \left(-\frac{2}{3},\frac{1}{3},-\frac{2}{3}\right)^T = d_1$ & $\left(-\frac{4}{\sqrt{21}},-\frac{\omega}{\sqrt{21}},\frac{2 \omega^2}{\sqrt{21}}\right)^T = (c_4 d_3)_{\mathbf{3}_S}$ \\
$ \left(-\frac{2}{3},\frac{\omega^2}{3},-\frac{2}{3} \omega\right)^T = d_2$ & $\left(-\frac{4}{\sqrt{21}},-\frac{1}{\sqrt{21}},\frac{2}{\sqrt{21}}\right)^T = (c_4 d_1)_{\mathbf{3}_S}$ \\
$ \left(-\frac{2}{3},\frac{\omega^2}{3},-\frac{2}{3} \omega\right)^T = d_2$ & $\left(-\frac{4}{\sqrt{21}},-\frac{\omega}{\sqrt{21}},\frac{2\omega^2}{\sqrt{21}}\right)^T = (c_4 d_3)_{\mathbf{3}_S}$ \\
$ \left(-\frac{2}{3},-\frac{2\omega^2}{3},\frac{\omega}{3}\right)^T = d_2^*$ & $\left(-\frac{4}{\sqrt{21}},\frac{2}{\sqrt{21}},-\frac{1}{\sqrt{21}}\right)^T = (c_4 d_1^*)_{\mathbf{3}_S}$ \\
$ \left(-\frac{2}{3},-\frac{2\omega^2}{3},\frac{\omega}{3}\right)^T = d_2^*$ & $\left(-\frac{4}{\sqrt{21}},\frac{2\omega}{\sqrt{21}},-\frac{\omega^2}{\sqrt{21}}\right)^T = (c_4 d_3^*)_{\mathbf{3}_S}$ \\[2mm]
\end{longtable}

\begin{longtable}[h!]{p{6.8cm}p{10cm}}
\caption{ The two EAs compatible with data in $3\sigma$ ranges in the inverted mass ordering. Each EA as a contraction of two flavon VEVs (not unique) is also shown.} \label{tab:EAs_IH}\\
\hline\hline

$V_1$ & $V_2$ \\\hline\\[-4mm] \endhead 
 $ \left(-\frac{1}{\sqrt{6}},-\frac{1}{\sqrt{6}},\sqrt{\frac{2}{3}}\right)^T = (a_1 d_1)_{\mathbf{3}_S} $ & $ \left(\frac{1-i}{\sqrt{6}},\frac{\left(3 \sqrt{2}-(1+2 i) \sqrt{6}\right)}{12},-\frac{\sqrt{3} (1+2 i)+3}{6 \sqrt{2}}\right)^T = (b_1^* c_2)_{\mathbf{3}_S} $ \\
 $ \left(-\frac{1}{\sqrt{6}},\sqrt{\frac{2}{3}},-\frac{1}{\sqrt{6}}\right)^T = (a_1 d_1^*)_{\mathbf{3}_S} $ & $ \left(\frac{1+i}{\sqrt{6}},\frac{-\sqrt{3} (1-2 i)-3}{6 \sqrt{2}},\frac{3-(1-2 i) \sqrt{3}}{6 \sqrt{2}}\right)^T = (b_1 c_2)_{\mathbf{3}_S} $ \\
 $ \left(-\frac{1}{\sqrt{6}},-\frac{\omega^2}{\sqrt{6}},\sqrt{\frac{2}{3}} \omega\right)^T = (a_2 d_2)_{\mathbf{3}_S} $ & $ \left(\frac{1+i}{\sqrt{6}},\frac{(3+3 i)+\sqrt{3} (1-5 i)}{12 \sqrt{2}},\frac{(-3-3 i)+\sqrt{3} (1-5 i)}{12 \sqrt{2}}\right)^T = (b_2^* c_1)_{\mathbf{3}_S} $ \\
 $ \left(-\frac{1}{\sqrt{6}},\sqrt{\frac{2}{3}} \omega^2,-\frac{\omega}{\sqrt{6}}\right)^T = (a_2 d_2^*)_{\mathbf{3}_S} $ & $ \left(\frac{1-i}{\sqrt{6}},\frac{(-3+3 i)+\sqrt{3} (1+5 i)}{12 \sqrt{2}},\frac{\sqrt{3} (1+5 i)+(3-3 i)}{12 \sqrt{2}}\right)^T = (b_2 c_1)_{\mathbf{3}_S} $ \\
 $ \left(-\frac{1}{\sqrt{6}},-\frac{\omega}{\sqrt{6}},\sqrt{\frac{2}{3}} \omega^2\right)^T = (a_3 d_3)_{\mathbf{3}_S} $ & $ \left(\frac{1-i}{\sqrt{6}},\frac{\sqrt{3} (1+5 i)+(3-3 i)}{12 \sqrt{2}},\frac{(-3+3 i)+\sqrt{3} (1+5 i)}{12 \sqrt{2}}\right)^T = (b_3^* c_1)_{\mathbf{3}_S} $ \\
 $ \left(-\frac{1}{\sqrt{6}},\sqrt{\frac{2}{3}} \omega,-\frac{\omega^2}{\sqrt{6}}\right)^T = (a_3 d_3^*)_{\mathbf{3}_S} $ & $ \left(\frac{1+i}{\sqrt{6}},\frac{(-3-3 i)+\sqrt{3} (1-5 i)}{12 \sqrt{2}},\frac{(3+3 i)+\sqrt{3} (1-5 i)}{12 \sqrt{2}}\right)^T = (b_3 c_1)_{\mathbf{3}_S} $ \\[2mm]\hline\hline
\end{longtable}

\end{document}